Ya-Wen Tsai,[1] Yao-Ting Wang,[2,3,5] Emanuele Galiffi,[3] Andrea Alù,[3,4] and Ta-Jen Yen[1,6]

[1] Department of Material Sciences and Engineering, National Tsing Hua University, Hsinchu 300, Taiwan.
[2] Department of Mathematics, Imperial College London, London SW7 2AZ, UK
[3] Photonics Initiative, Advanced Science Research Center, City University of New York, New York, NY 10031, USA
[4] Physics Program, Graduate Center of the City University of New York, New York, NY 10016, USA
[5*] ywang5@gc.cuny.edu
[6*] tjyen@mx.nthu.edu.tw

# Surface-Wave Coupling in Double Floquet Sheets Supporting Phased Temporal Wood Anomalies

**Abstract:** We investigate symmetry-selective surface-mode excitation in a general periodically time-modulated double-layer system, where the modulation of the two layers has a constant phase difference. By deriving a semi-analytic transfer matrix formalism of a Drude-dispersive double-layer structure with periodic time-modulation, we calculate the scattering amplitudes and the corresponding transmission coefficient. Our results show that the phase-difference between the modulation of the two sheets plays an essential role in significantly enhancing and selectively exciting either the even or odd surface mode with high efficiency. We verify our calculations with full-wave time-domain simulations, showing that efficient switching between the surface-wave excitation of the two distinct modal channels can be achieved, even under illumination from a single off-resonant Gaussian pulse, by controlling the phase difference between the two modulations. Our results pave the way towards ultrafast, symmetry-selective mode excitation and switching via temporal modulation.

## 1 Introduction

Confining and manipulating light at the nano-scale constitutes a decades-long scientific and technological challenge, with applications ranging from sensing and spectroscopy to imaging, communications, the optical probing of quantum materials and quantum optics. One fast-growing family of electromagnetic resonances capable of confining light to subwavelength volumes is represented by surface waves (SWs), travelling-wave excitations propagating along the interface between different materials, while decaying exponentially away from it. The first type of SWs exploited in nanophotonics is the surface plasmon polariton, which originates from the coupling between an impinging electromagnetic field and oscillations of an electron plasma [1]. More recently, a plethora of new polaritonic materials and excitations have been discovered, often featuring exotic dispersion relations, such as hyperbolic [2], ghost [3] and shear [4] phonon polaritons in anisotropic media.

On the other hand, the combined rise of ultra-thin, highly tunable layered materials and polaritonics has drawn significant attention to the opportunities stemming from leveraging the strong light-matter interactions enabled by these materials, and the resulting nonlinearities, to achieve temporal control of light [5,6] and matter [7]. This mechanism has led to a surge of interest in time-varying electromagnetic systems, which hold the promise to endow the fields of photonics, metamaterials and condensed matter physics with an additional degree of freedom for tailoring wave-matter interactions. Light-matter dynamics in time-modulated systems have been demonstrated to host a wealth of exotic wave phenomena, including parametric amplification [8,9], negative refraction [10], time-refraction [11] and frequency shifting [12-14], photon acceleration [15] and the engineering of synthetic frequency dimensions [16]. In addition, a rising number of theoretical proposals have been studied, such as synthetic motion [17-18], temporal topological edge states [19], nonreciprocity [20-21] and space-time metamaterials [22-24], temporal aiming [25] and antireflection coatings [26], nonlocality [27], Floquet topology [28-31] and chirally selective amplification [32], among many others [33].

One particular avenue that was recently proposed in this context is the one of dynamical Wood anomalies [34]: resonant, efficient SW-excitation schemes that form the analogue of spatial grating couplers to excite SWs from the far-field, by exploiting a periodic temporal modulation in the electromagnetic response of a spatially homogeneous material, as it may be realized via nonlinearities in a polaritonic medium such as graphene [35]. In this work, we extend the concept of dynamical Wood anomalies by introducing a second Floquet sheet, whose periodic time-modulation differs from the first one by a constant phase delay. While the additional sheet introduces a second surface mode with opposite symmetry, the time-grating can only couple radiation efficiently to one specific mode for a given interlayer gap, due to symmetry mismatch. However, by tuning the constant modulation phase between the two Floquet sheets, symmetry is broken and waves impinging from the far-field can be made to selectively couple to surface modes with a specific symmetry, with high efficiency. We provide semi-analytical transfer matrix solutions to carry out the scattering problem, and verify our theoretical findings via both frequency domain and time-domain full-wave numerical simulations, concluding with numerical experiments where we selectively excite even or odd modes with a single pulse, by solely changing the interlayer phase. Our results introduce new opportunities stemming from ultrafast, dynamical switching to control the coupling to different surface mode channels on a time-modulated surface, or metasurface, paving the way for new directions in the context of dynamical Wood anomalies and polaritonic Floquet-surfaces.

## 2 Principle

We begin by considering two temporally modulated sheets that divide the space into three regions, with dielectric constants $\varepsilon_1$, $\varepsilon_2$, and $\varepsilon_3$. The schematic configuration of the proposed structure is depicted in Fig. 1(a). The current density $J(t)$ on the temporal sheets can be described by the Drude dynamic equation [34]

$$\frac{dJ(t)}{dt} + \gamma J(t) = W_D(t) E_x(x,t) \tag{1}$$

where $\gamma$ is a phenomenological dissipation rate, $W_D(t)$ is the Drude weight, proportional to the density of free charge carriers at time $t$, and $E_x(x,t)$ is the in-plane component of the electric field for the TM-polarized waves considered. We assume that the Drude weight $W_D(t)$ is periodically modulated in time as $W_D(t) = W_{D0}\left[1 + 2\alpha \cos\left(\Omega t + \phi_j\right)\right]$, where $\alpha$, $\Omega$, and $\phi_j$ are the strength, frequency and phase of the modulation on the respective Floquet sheets, as shown in Fig. 1(a).

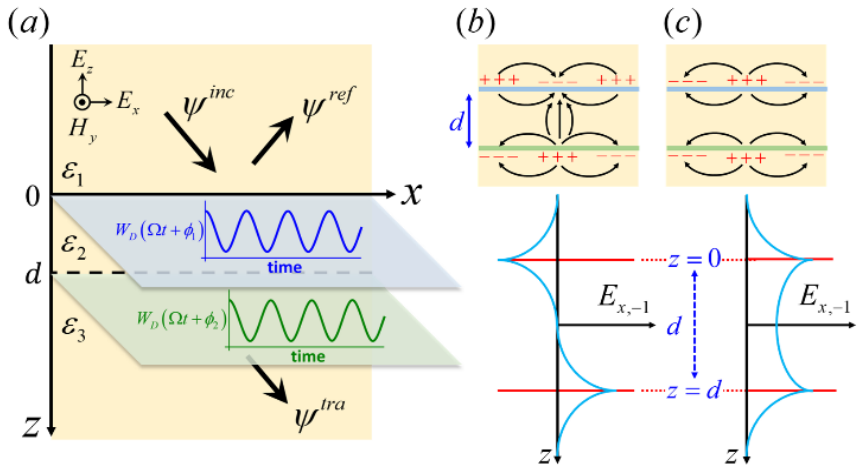

**Fig. 1.** (a) Schematic of a double-layer system with an interlayer spacing d along the z-axis. The temporal modulation of both sheets enters the system via the Drude weight $W_D$. The dielectric sub-domains are characterized by dielectric constants $\varepsilon_1$, $\varepsilon_2$, and $\varepsilon_3$. (b,c) Symmetry of the x-component of the electric fields (for the $n = -1$ harmonic channel, $E_{x,-1}$), for the anti-symmetric (b) and symmetric (c) surface modes.

We start by considering a TM polarized wave $\left(E_x, E_z, H_y\right)$ with in-plane electric field:

$$E_{jx}\left(x,z,t\right)=\sum_n\left(E_{jx,n}^+e^{ik_{z,n}^{(j)}z}+E_{jx,n}^-e^{-ik_{z,n}^{(j)}z}\right)e^{i\left[k_xx-\left(\omega+n\Omega\right)t\right]}\tag{2}$$

where $E_{jx,n}^\pm$ are the complex amplitudes of the different forward (backward) propagating Floquet harmonics, $k_x$ is the in-plane wave-vector, and $k_{z,n}^{(j)}=\sqrt{\varepsilon_j\left(\omega+n\Omega\right)^2/c^2-k_x^2}$ is the component of the wave-vector perpendicular to the interfaces. The factors $n$, $\omega$ and $c$ in $k_{jz}$ denote the order of the Floquet harmonic, angular frequency, and the speed of light in vacuum, respectively, while the indices $j=1,2,3$ label the regions with different dielectric constants. From Maxwell's equations, the magnetic field components $H_{jy,n}^\pm$ is

$$H_{jy,n}^\pm=\pm\frac{\left(\omega+n\Omega\right)\varepsilon_0\varepsilon_j}{k_{z,n}^{(j)}}E_{jx,n}^\pm.\tag{3}$$

To determine the amplitudes, we enforce the continuity of the in-plane electric fields $E_{jx}$ and discontinuity of the magnetic fields $H_{jy}$ by the surface current $J\left(t\right)=\sum_n J_n e^{i\left[k_xx-\left(\omega+n\Omega\right)t\right]}$, where

$$J_n=W_{D0}\frac{E_n+\alpha\left(E_{n+1}e^{i\phi}+E_{n-1}e^{-i\phi}\right)}{\gamma-i\left(\omega+n\Omega\right)}.\tag{4}$$

Thus, in frequency space, applying the boundary conditions at the two current sheets, leaves us with a $2n\times2n$ transfer matrix $\mathbf{M}$ for the double Floquet sheet (DFS) system,

$$\left[\cdots,E_{1x,n-1},E_{1x,n},E_{1x,n+1},\cdots\right]^T=$$
$$\mathbf{M}_{2n\times2n}\left[\cdots,E_{3x,n-1},E_{3x,n},E_{3x,n+1},\cdots\right]^T,\tag{5}$$

where $\mathbf{M}_{2n\times2n}=\left[\mathbf{D}_{12}\mathbf{P}_2\left(d\right)\mathbf{D}_{23}\right]_{2n\times2n}$. The explicit form of the matching matrices $\mathbf{D}_{12}$ and $\mathbf{D}_{23}$, and the propagation matrix $\mathbf{P}_2\left(d\right)$ is given in Supplement 1. Since convergence was checked against the 3-, 5- and 7-harmonics cases, shown in Supplement 1, we simplify the problem by truncating the transmitted matrix to the first three Floquet modes as

$$\begin{pmatrix}0\\E_{1x,0}^+\\0\end{pmatrix}=M_{3\times3}^{tra}\begin{pmatrix}E_{3x,1}^+\\E_{3x,0}^+\\E_{3x,-1}^+\end{pmatrix}=\left[D_{1,2}P_2\left(d\right)D_{2,3}\right]_{3\times3}\begin{pmatrix}E_{3x,1}^+\\E_{3x,0}^+\\E_{3x,-1}^+\end{pmatrix}\tag{6}$$

With the reduced transfer matrix in Eq. (6), the approximate solutions of the transmitted amplitudes with and without time-modulation can then be obtained. Assuming that the incoming wave is incident from the top [see Fig. 1(a)] upon the DFS with transmission coefficient denoted by T, it can be shown that the transmission coefficient for the fundamental harmonic is determined in terms of the elements of $M_{3\times3}^{tra}$ as

$$\mathrm{T}=\frac{E_{3x,0}^+}{E_{1x,0}^+}=\frac{m_{11}m_{33}-m_{13}m_{31}}{\det\left(M_{3\times3}^{tra}\right)}\tag{7}$$

In addition, the dispersion relation of the surface waves in the unmodulated structure can also be obtained when the matrix $M_{3\times3}^{tra}$ satisfies the condition $\det\left(M_{3\times3}^{tra}\right)=0$. There are two SW solutions, corresponding to anti-symmetric and symmetric modes, as shown in Figs. 1(b) and 1(c). In the anti-symmetric mode, in-plane charge oscillations along the two sheets are of opposite sign [Fig. 1(b)], whereas they are of the same sign for the symmetric mode [Fig. 1(c)]. For the sake of generality, in the following we introduce the dimensionless parameters $\tilde\omega=\omega/\left(W_{D0}Z_0\right)$, $\tilde k_x=k_xc/\left(W_{D0}Z_0\right)$, and $\tilde d=d\left(W_{D0}Z_0\right)/c$, where $Z_0=\sqrt{\mu_0/\varepsilon_0}$ is the impedance of free space.

# 3  Double Floquet sheets

In order to clearly show the mechanism of temporal Wood anomalies on our DFS, we first solve the dispersion relation for the unmodulated structure at a fixed interlayer distance $\tilde d=2$, shown in Fig. 2(a). The two surface modes, denoted by blue features outside the light line, have resonant frequencies $\tilde\omega_{SW-}\approx0.456$ for the antisymmetric mode and $\tilde{\omega}_{SW_+} \approx 0.576$ for the symmetric one for given incoming wavevector $\tilde{k}_x = 0.75$ (white vertical dashed line). As both SW modes lie outside the light cone, they cannot be excited directly by propagating waves due to momentum (or energy) mismatch. In order to overcome this limitation, we exploit the temporal modulation to induce transitions between the incoming propagating waves and the SW modes across the light line, a concept recently introduced in the context of temporal Wood anomalies [34]. Thus, waves with frequency $\tilde{\omega}_0$ and wavevector $\tilde{k}_x$ can be efficiently coupled to surface modes when the modulation frequency $\tilde{\Omega}$ matches the frequency gap $\Delta\tilde{\omega}_\pm$ between incident waves and surface modes. For given modulation frequency $\tilde{\Omega} \approx 0.6$, Fig. 2(b) shows the transmittance spectra of the DFS system with and without time modulation, ($\alpha = 0.15$ and $\alpha = 0$ respectively). The time-modulated case reveals two clear transmittance dips at $\tilde{\omega}_{0-} \approx \tilde{\omega}_{SW-} + \tilde{\Omega} \approx 1.056$ and $\tilde{\omega}_{0+} \approx \tilde{\omega}_{SW+} + \tilde{\Omega} \approx 1.176$, indicating that a large amount of the incident energy is trapped into surface waves. This effect is reflected in the fact that the harmonic of order $n = -1$ strongly dominates the scattering process over the generation of other harmonics, a typical trait of Wood scattering anomalies.

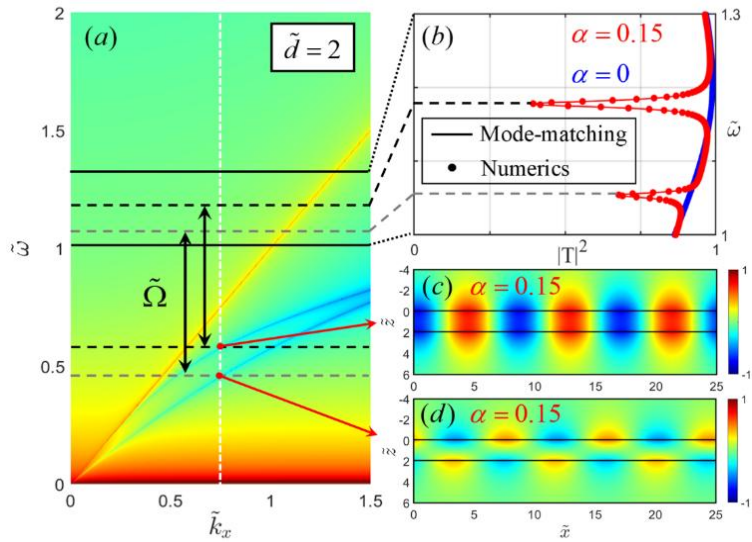

**Fig. 2.** Dispersion relation of surface modes at a double-sheet system, calculated using $M$. The black double arrows indicate the excitation mechanism with time-gratings at the modulation frequency $\tilde{\Omega}$. (b) Transmittance spectra for two temporal sheets with and without modulation $\alpha = 0.15$ and $0$ at fixed in-plane wavevector $\tilde{k}_x = 0.75$ [white vertical dashed line in (a)]. Here we used $\tilde{d} = 2$, $\varepsilon_1 = \varepsilon_2 = \varepsilon_s = 1$, $\tilde{\Omega} \approx 0.6$, and $\tilde{\gamma} \approx 0.01$. Solid line: Mode-matching result. Dots: Numerical result from COMSOL. (c,d) Electric field $\tilde{E}_{x,-1}$ distributions of the symmetric and anti-symmetric surface modes under time-modulation of its Drude weight with modulation frequency $\tilde{\Omega} \approx 0.6$. Symmetric (even) mode at frequency $\tilde{\omega}_{0+} \approx 1.176$ and anti-symmetric (odd) mode at frequency $\tilde{\omega}_{0-} \approx 1.056$ are shown.

To characterize the two SW modes, we have simulated the DFS system in the x-z plane with FEM-based software COMSOL 5.6. With time modulation, Figs. 2(c) and 2(d) show the $x$ component of the electric fields in the $n = -1$ harmonic channel, i.e., $\tilde{E}_{x,-1}$, demonstrating the opposite symmetry of the respective modes. However, the two transmittance dips in Fig. 2(b) reveal that 5% and 70% of transmitted power is converted to the respective SWs. Thus, whilst significant, the SW excitation efficiency is weak, especially for the anti-symmetric mode, which is almost dark due to poor symmetry matching with the incoming wave. In the next section, we show that a constant phase difference between the modulation of the two sheets in the DFS system can lead to strong SW conversion efficiency, up to 80%. Remarkably, this SW excitation can be selectively achieved for the symmetric or anti-symmetric mode by varying the relative phase.

# 4 Phased double Floquet sheets

Conventionally, SW excitation can normally be improved for either the symmetric or antisymmetric mode by tuning the distance between the sheets, which however necessarily makes the other mode darker. A temporal Wood anomaly can overcome this limitation when a constant phase shift is introduced between the two time-gratings, enabling dynamic switching between coupling to the two modes. In our scenario, Fig. 3(a) shows the transmittance spectra for the fundamental harmonic of the fields scattered off the DFS structure when the two Floquet sheets are modulated in phase, as a contour plot in terms of frequency $\tilde{\omega}$ and interlayer distance $\tilde{d}$. As $\tilde{d}$ is reduced, the frequency spacing between the two modes increases as a result of the stronger interaction between the near fields. Simultaneously, the two modes alternate between being dark and bright, as a result of the change in phase accumulation of the propagating waves between the two sheets, which can only resonantly excite one of the two modes efficiently for a given interlayer gap $\tilde{d}$.

In the scenario where the two time-gratings differ by a constant relative phase $\Delta\phi$, in Fig. 3(b) we show the transmittance spectrum $\left| T\left(\tilde{\omega}, \Delta\phi\right)\right|^{2}$, where we vary the phase difference $\Delta\phi$ between the two sinusoidal modulations from 0 to $2\pi$ and the incident frequency $\tilde{\omega}$ from 1 to 1.25, for a fixed incoming wavevector $\tilde{k}_{x}=0.75$ and interlayer distance $\tilde{d}=2$. The results show how proper tuning of the constant phase difference between the two sheets can achieve strong coupling efficiency to either one of the surface modes. Fig. 3(c) depicts the transmittance $|T|^{2}$ of odd and even modes, which correspond to the black and magenta vertical-dashed lines in Fig. 3(b), as a function of the phase difference $\Delta\phi$. The minimum values of the transmittance of the first and second modes are 0.168 and 0.198 at $\Delta\phi\approx1.4\pi$ and $\Delta\phi\approx0.4\pi$, respectively. Compared to the results in Fig. 2(b), the SW excitation efficiencies reach values larger than 80%, controlled by the relative phase difference between the respective temporal modulations. By comparison with the time modulation results in Fig. 3(a), Fig. 3(b) shows that a symmetry selective effect can be achieved by controlling the relative phase shift, which may be preferable to varying $\tilde{d}$, as it can be switched dynamically without changing the structure. In addition, such phase difference can be used to select the symmetry of the surface mode, as detailed in Fig. 3(c). As the phase difference between the two modulation sheets changes by $\pi$, the excitation of the symmetric mode can be switched to the anti-symmetric mode, and vice versa. This comes as a result of the extra phase accumulated by the waves as they scatter within the DFS. This phase shift, however, is no longer provided by propagation, but by the very interaction with the sheets. The corresponding $\tilde{E}_{x,-1}$ field distributions of the symmetric and anti-symmetric surface modes in the x-z plane are shown in Figs. 3(d-g), with the same normalization factor and color mapping. At the resonant frequency $\tilde{\omega}_{0-}$, (panel d,f), the antisymmetric surface mode can be switched between bright and dark by shifting $\Delta\phi$ by $\pi$. Conversely, at $\tilde{\omega}_{0+}$ the symmetric mode can be excited with a very high or very low efficiency by choosing $\Delta\phi$ adequately. This phase-enabled symmetry selection is completely reconfigurable, in principle at very fast timescales, enabling an additional orthogonal coupling channel for the trapping of waves from the continuum.

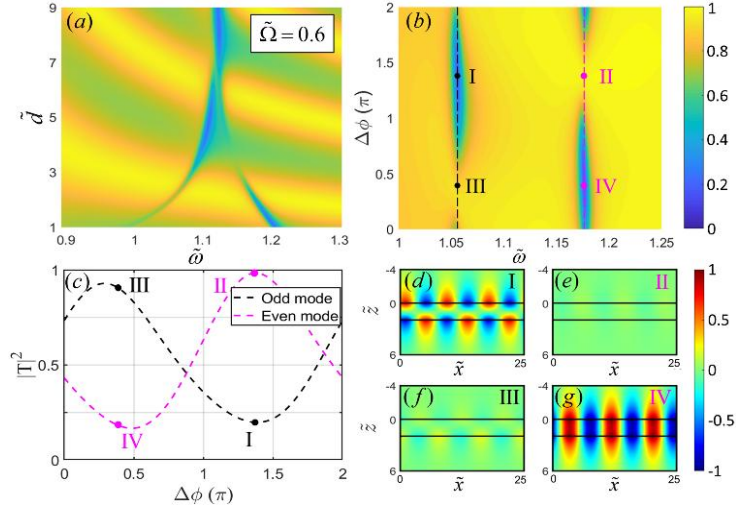

**Fig. 3.** (a) Transmittance spectra of the DFS structure without phase modulation as a function of incident frequency $\tilde{\omega}$ and interlayer distance $\tilde{d}$. (b) Contour plots of the transmittance spectra as a function of incident frequency $\tilde{\omega}$ and phase difference $\Delta\phi$. The positions of the black and magenta vertical-dashed lines represent the two resonance frequencies in Fig.2(b), and the four dots highlight the maximum and minimum values of the transmittance, labeled as I, II, III, IV, on the black and magenta vertical-dashed lines. (c) Transmittance through the DFS for the cases of efficient excitation of the odd (anti-symmetric) and even (symmetric) modes at the respective resonant frequencies (corresponding to the black and magenta vertical-dashed lines in (b)), as a function of the phase difference $\Delta\phi$. (d-g) Amplitude distributions of electric field $\tilde{E}_{x,-1}$ corresponding to the four dots marked in (b).

The switching capability of our DFS system is best exploited if just the phase shift can be leveraged to selectively excite a specific mode given the same input wave. In order to investigate this avenue, we design three time-domain numerical simulations with Gaussian pulses of different carrier frequencies [see Supplementary 1 for details on simulation settings]. Fig. 4 shows a snapshot of the in-plane components of the electric field $\tilde{E}_p$ at time $\tilde{t} \approx 186$, after the scattering has occurred. In Figs. 4(a) and 4(b), we consider two narrowband Gaussian pulses with different carrier frequencies resonant with the two respective Wood anomalies, i.e., $\tilde{\omega}_{0,-} \approx 1.056$ and $\tilde{\omega}_{0,+} \approx 1.176$. The DFS structures are tilted to angles $\theta \approx 45°$ and $\theta \approx 39.6°$ relative to the pulse direction, in order for their carrier wave to match the in-plane momentum $\tilde{k}_x = 0.75$. When the pulse interacts with the DFS, it efficiently excites the corresponding SW mode at that frequency [Figs. 4(a-b)], which propagates along the DFSs, as expected with a coupling efficiency that can be maximized via the constant phase difference between the modulation at the two sheets. [Full animations available in the supplementary material].

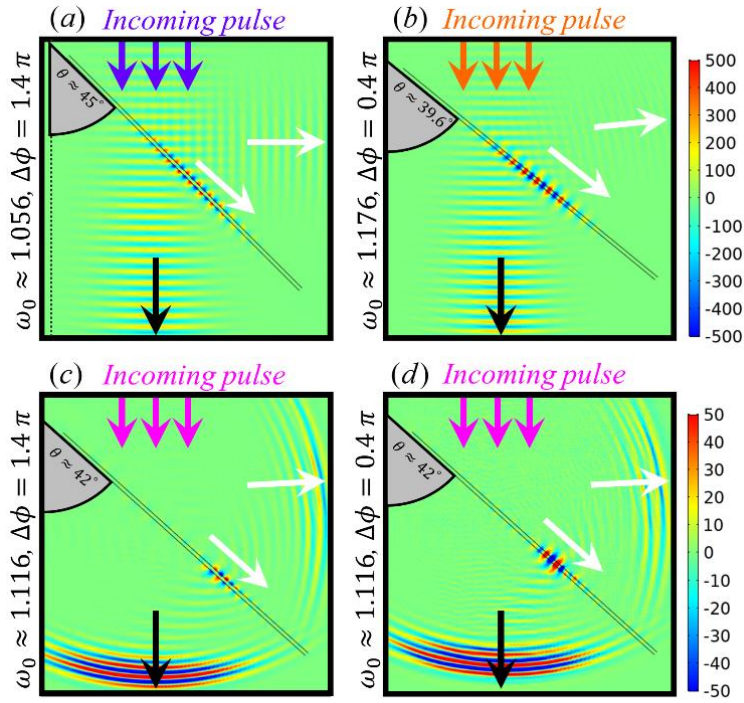

**Fig. 4.** Temporal simulations of phased dynamical Wood anomalies predicted in Fig. 3. (a-b) On-resonance pulses with carrier frequencies $\bar{\omega}_{0,-} \approx 1.056$ (purple arrows) and $\bar{\omega}_{0,+} \approx 1.176$ (orange arrows) impinge on the DFS $\alpha = 0.15$ with phase differences (a) $\Delta\phi \approx 1.4\pi$ and (b) $\Delta\phi \approx 0.4\pi$, individually optimized to excite anti-symmetric and symmetric SWs in the respective cases with high efficiency. (c,d) An off-resonant pulse with carrier frequency $\bar{\omega}_c \approx 1.116$ (pink arrows) impinges on the time-phased DFS $\alpha = 0.15$ with $\Delta\phi \approx 1.4\pi$ and $\Delta\phi \approx 0.4\pi$, efficiently exciting SWs of either symmetry based uniquely on the modulation phase $\Delta\phi$.

To achieve symmetry-selection at the same incident angle and frequencies, we design a Gaussian pulse whose bandwidth covers the two resonance frequencies [pink curve in Fig. S3]. The off-resonance carrier frequency is chosen in the middle of the two resonance frequencies, i.e., $\bar{\omega}_{0,mid} \approx 1.116$, and the full width at half maximum of the pulse is FWHM $\approx 0.34$, 3 times wider than the difference between the two resonant frequencies, such that the pulse intensity between the two resonant frequencies is above 80%. With this Gaussian pulse, Figs. 4(c) and 4(d) reveal that SWs with a specific symmetry can be excited with high efficiency simply by varying the relative modulation phase, without changing the carrier frequency of the Gaussian pulse. This symmetry-selective coupling provides an effective way of realizing optimal far-field excitation of a desired SW channel on an unstructured interface, relying solely on the temporal modulation to enable the coupling, and drastically reconfigurable through the phase delay $\Delta\phi$.

We envision possible implementations of a DFS device in high-quality graphene, either via all-optical modulation or electrical bias. All-optical modulation refers to the capability of graphene to undergo ultrafast modulations of its graphene carrier density under infrared pumping [37,38]. In addition, electrical bias refers to the ultrafast response of graphene to incident electrostatic fields, which has been previously used to realize graphene modulations [39]. In order to match our parameters with realistic values, experimentally achieved for carrier modulation speeds and graphene plasmon lifetimes, we can choose to work in a lower frequency regime, where the SW dispersions are closer to the light line, as detailed in Fig. S4 of the Supplementary material. In Fig. 5, we demonstrate the reflection spectrum for incoming terahertz waves with in-plane wavevector $k_x = 8.87$ rad/mm impinging on the DFS device as $d \approx 0.17$ mm. Given a realistic modulation frequency $f_m \approx 0.16$ THz, the two in-coupled photon frequencies are $f_{0,-} = 0.515$ THz and $f_{0,+} = 0.57$ THz, corresponding to odd and even surface modes. Here we assume a modulation amplitude $\alpha = 0.15$, and demonstrate the effect of three different loss rates $\gamma = 0.3$, 0.6 and 1.2 THz. Clearly, the onset of losses plays a role both in the coupling to the two eigenmodes, but also in the selectivity achievable, due to the loss of orthogonality between the two eigenmodes with opposite parity. Furthermore, due to its lower frequency, the lifetime of the antisymmetric mode is more affected by the loss, so that the corresponding excitation signal becomes hardly distinguishable for loss rates above 1 THz, a common problem of acoustic plasmons in graphene. However, for reasonably high-quality samples, it is possible to excite preferentially one of the two SW modes, while suppressing the other one by simply tuning the phase difference $\Delta\phi$ between two graphene layers by $\pi$ radians. Thus, although both reflection signals originating from the SW coupling drop as the loss rate increases, symmetry-selective coupling can still be achieved under a reasonable loss rate. Finally, this model has further opportunities for experimental implementations in the RF by variable capacitors, whose effective impedance can be modulated at GHz speeds, and the designed waves can be tailored by adequately structuring the metasurfaces.

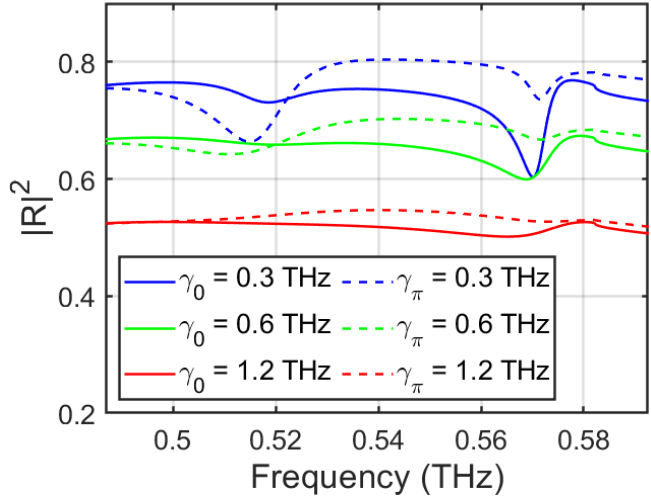

**Fig. 5.** Numerical calculations of the reflection spectrum of phased double-Floquet-graphene layers at $d \approx 0.17$ mm for different values of loss rates $\gamma_{\Delta\phi} = ev_F^2 / mE_F$ from 0.3 THz (blue), 0.6 THz (green), to 1.2 THz (red) based on a conservative electron mobility $m = 100 \times 10^3$, $50 \times 10^3$, $25 \times 10^3$ cm²/ V·s [40]. Solid and dashed lines correspond to $\Delta\phi \approx 0$ and $\Delta\phi \approx \pi$, corresponding to the preferential even and odd mode excitations respectively. The parameters are $E_{F,0} = 0.3$ eV, $W_{D,0} = e^2 E_{F,0} / \pi\hbar^2 = 0.035$ THz, $f_m = \Omega/2\pi \approx 0.16$ THz, and $v_F = 9.5 \times 10^7$ cm/s [34], which correspond to the Fermi level, Drude weight, modulation frequency, and Fermi velocity of the charge carriers, respectively.

# 5 Concluding remarks

To conclude, we have investigated double time-modulated Floquet sheets to selectively excite symmetric and anti-symmetric SW modes at flat interfaces. By the means of a general, efficient transfer-matrix model, we showed how symmetry-selective temporal Wood anomalies, capable of enhancing and efficiently switching between the excitation of even and odd surface modes, can be engineered by introducing a constant phase difference between the temporal modulation of the two conductive layers, simultaneously suppressing the undesired mode. Furthermore, we used time-domain Gaussian pulses impinging on the DFS in transient FEM simulations to demonstrate how the same impinging signal can be efficiently coupled to either symmetric or anti-symmetric mode by controlling the relative modulation phase. In practice, implementations are viable either within the THz range, where the switching effect can be realized by either all-optical or electrical pumping, or at microwaves, by exploiting active metasurfaces supporting designer SWs, such as spoof plasmons.

Our results open a new avenue for ultrafast and fully reconfigurable optoelectronic switching, with further opportunities stretching from the engineering of geometric phases via more complex dynamical phase difference and modulation schemes with additional layers, to further applications of these concepts for the design of broadband active absorbers and amplifying devices.

**Research funding:** Ministry of Science and Technology (MOST 110-2221-E-007-051-MY3, MOST 110-2218-E-007 -055-MBK, MOST 111-2923-E-007-007-MY2).

**Acknowledgments:** This work was supported by grants from the Ministry of Science and Technology (MOST 110-2221-E-007-051-MY3, MOST 110-2218-E-007 -055-MBK, MOST 111-2923-E-007-007-MY2). E.G. was supported via a Junior Fellowship of the Simons Society of Fellows (855344,EG).

**Conflict of interest statement:** The authors declare that there are no conflicts of interest related to this article.

**Supplemental document.** See Supplement 1 for supporting content.